\documentclass[runningheads]{llncs}
\usepackage{graphicx}
\usepackage{amssymb,amsmath}
\usepackage{tikz}
\begin{document}

\title{Self-Supervised Semantic Guidance For Structure Preserving Stain Normalization}

\titlerunning{Structure Preserving Stain Normalization of Histopathology Images}

\author{Dwarikanath Mahapatra \inst{1} \and
Behzad Bozorgtabar \inst{2,3,4} \and Jean-Philippe Thiran \inst{2,3,4} \and
Ling Shao \inst{1,5} }
 
 % index{Mahapatra, Dwarikanath}
 % index{Bozorgtabar, Behzad}
 % index{Thiran, Jean-Philippe}
 % index{Shao, Ling}
 
\authorrunning{D. Mahapatra et al.}

\institute{Inception Institute of Artificial Intelligence, Abu Dhabi, UAE \email{\{dwarikanath.mahapatra,ling.shao\}@inceptioniai.org} \and{Signal Processing Laboratory 5, EPFL, Lausanne, Switzerland \email{\{behzad.bozorgtabar,jean-philippe.thiran\}@epfl.ch} 
 \and {Department of Radiology, Lausanne University Hospital, Lausanne, Switzerland}
 \and {Center of Biomedical Imaging, Lausanne, Switzerland}
\and Mohamed bin Zayed University of Artificial Intelligence, Abu Dhabi, UAE}}
\maketitle              % typeset the header of the contribution
%
%%%%%%%%% ABSTRACT
\begin{abstract}
Although generative adversarial network (GAN) based style transfer is state of the art in histopathology color-stain normalization, they do not explicitly integrate structural information of tissues. We propose a self-supervised approach to incorporate semantic guidance into a GAN based stain normalization framework and preserve detailed structural information. Our method does not require manual segmentation maps which is a significant advantage over existing methods. We integrate semantic information at different layers between a pre-trained semantic network and the stain color normalization network. The proposed scheme outperforms other color normalization methods leading to better classification and segmentation performance.
  
   
\end{abstract}
%

%%%%%%%%% BODY TEXT

\section{Introduction}
\label{sec:intro}
Increased digitization of pathology slides has enhanced the importance of digital histopathology in the medical imaging community \cite{bozorgtabar2020computational,Mahapatra_PR2020,ZGe_MTA2019,Behzad_PR2020,Mahapatra_CVIU2019,Mahapatra_CMIG2019,Mahapatra_LME_PR2017}. Staining is an important part of pathological tissue preparation where, e.g., Hematoxylin and Eosin dyes alter the intensity of tissue elements - nuclei turn dark purple while other structures become pink. Tissue structures become distinguishable, facilitating manual or automated analysis.

The color variation of the same structure is observed due to differences in staining protocols from different centers, different dye manufacturers, and scanner characteristics. Consequently, this leads to inconsistent diagnosis and limits the efficacy of automated methods. Hence there is a need for stain color normalization to have a uniform appearance of dye-stained regions. We propose to integrate self-supervised semantic guidance with GANs for better structure preservation after stain normalization.

Two widely explored categories for stain normalization methods are color matching \cite{Reinhard,Zilly_CMIG_2016,Mahapatra_SSLAL_CD_CMPB,Mahapatra_SSLAL_Pro_JMI,Mahapatra_LME_CVIU,LiTMI_2015,MahapatraJDI_Cardiac_FSL,Mahapatra_JSTSP2014}, and stain-separation \cite{Machenko,SG9,MahapatraTIP_RF2014,MahapatraTBME_Pro2014,MahapatraTMI_CD2013,MahapatraJDICD2013,MahapatraJDIMutCont2013,MahapatraJDIGCSP2013,MahapatraJDIJSGR2013}. Since these methods rely on template images, it leads to mismatch and poor performance when the template is not representative of the dataset. 
The third category comprises machine learning approaches \cite{Zhou19_6,MahapatraJDISkull2012,MahapatraTIP2012,MahapatraTBME2011,MahapatraEURASIP2010,Mahapatra_CVPR2020,Kuanar_ICIP19,Bozorgtabar_ICCV19} which sub-divide an input image into multiple tissue regions using a sparse autoencoder, and independently normalize each region.  
Recent works solve stain normalization as a style-transfer problem using Generative adversarial networks(GANs) \cite{Taieb,Xing_MICCAI19,Mahapatra_ISBI19,MahapatraAL_MICCAI18,Mahapatra_MLMI18,Sedai_OMIA18,Sedai_MLMI18,MahapatraGAN_ISBI18}. GANs have found many applications in medical image analysis \cite{GAN_MI_Rev,GAN_MI_Rev2,Sedai_MICCAI17,Mahapatra_MICCAI17,Roy_ISBI17,Roy_DICTA16,Tennakoon_OMIA16,Sedai_OMIA16,Mahapatra_MLMI16} such as image super-resolution \cite{MahapatraMICCAI_ISR,mahapatra2019progressive,mahapatra2017retinal,rad2020benefitting}, registration \cite{Mahapatra_PR2020,Sedai_EMBC16,Mahapatra_EMBC16,Mahapatra_MLMI15_Optic,Mahapatra_MLMI15_Prostate,Mahapatra_OMIA15,MahapatraISBI15_Optic,MahapatraISBI15_JSGR}, segmentation \cite{ZhaoMic18,MahapatraAL_MICCAI18,MahapatraISBI15_CD,KuangAMM14,Mahapatra_ABD2014,Schuffler_ABD2014,MahapatraISBI_CD2014,MahapatraMICCAI_CD2013,Schuffler_ABD2013}, domain adaptation \cite{bozorgtabar2019syndemo,bozorgtabar2020exprada,bozorgtabar2019learn,bozorgtabar2019using,MahapatraProISBI13,MahapatraRVISBI13,MahapatraWssISBI13,MahapatraCDFssISBI13,MahapatraCDSPIE13,MahapatraABD12,MahapatraMLMI12,MahapatraSTACOM12} and data augmentation \cite{Mahapatra_CVPR2020,Mahapatra_CVIU2019,VosEMBC,MahapatraGRSPIE12,MahapatraMiccaiIAHBD11,MahapatraMiccai11,MahapatraMiccai10,MahapatraICIP10,MahapatraICDIP10a,MahapatraICDIP10b} to name a few. Unpaired Image-to-Image Translation with CycleGANs were used in \cite{StainGAN,MahapatraMiccai08,MahapatraISBI08,MahapatraICME08,MahapatraICBME08_Retrieve,MahapatraICBME08_Sal,MahapatraSPIE08,MahapatraICIT06,DART2020_Ar,CVPR2020_Ar} to facilitate style transfer across two domains. These methods do not require a reference image and achieve high visual agreement with images from the target domain. 
 Gupta et al. \cite{Gupta,sZoom_Ar,CVIU_Ar,AMD_OCT,GANReg1_Ar,PGAN_Ar,Haze_Ar,Xr_Ar,RegGan_Ar,ISR_Ar} leverage GAN based image-image translation for data augmentation of histopathology images, yielding an improvement in segmentation accuracy. Other variants include the use of prior latent variables and auxiliary networks \cite{Zanjani,LME_Ar,Misc,Health_p,Pat2,Pat3,Pat4,Pat5,Pat6,Pat7},  and auxiliary inputs \cite{ZhouMiccai19}.

Previous works have demonstrated the effectiveness of cycle GANs in stain normalization, thus eliminating the tedious task of selecting a reference stain. However, as pointed out in \cite{gadermayr2018miccai,Pat8,Pat9,Pat10,Pat11,Pat12,Pat13,Pat14,Pat15,Kuanar_AR1,Lie_AR2} shape outlines of translated objects may change which leads to sub-optimal performance. Gadermayr \cite{gadermayr2018miccai} used two different pipelines to overcome this pitfall. While their results are effective, the pipeline itself is tedious.
 Vahadane et al. \cite{Vahadane} propose a structure-preserving normalization method using non-negative matrix factorization but do not explicitly use semantic information. Lahiani et al. \cite{Lahiani} introduce a perceptual embedding loss to reducing tiling artifacts in reconstructed whole slide images (WSI).

Self-supervised learning requires formulating a proxy (or pretext) task, which can be solved on the same dataset and using the trained network to perform self-supervised tasks such as segmentation or depth estimation \cite{SemGuidICLR20}. 
 Some examples in the field of medical image analysis include divide-and-rule framework for survival analysis in colorectal cancer \cite{abbet2020divide}, surgical video re-colorization as a pretext task for surgical instrument segmentation \cite{BaiMic19_10}, rotation prediction for lung lobe segmentation and nodule detection \cite{BaiMic19_11} and use disease bounding box localization for cardiac MR image segmentation \cite{BaiMiccai19}.

% \textbf{Our Contributions:} 
\subsubsection{Contributions:}
Since medical image analysis influences diagnostic decisions, it is helpful to preserve information about more delicate structures for semantic guidance. The inclusion of segmentation information requires detailed annotations of the image, which is extremely cumbersome for WSIs.
Our primary contribution is a color stain normalization method that uses semantic guidance through self-supervised features. We build our model using cyclic GANs \cite{cycleGAN,Mahapatra_PR2020,Pat17,Pat16,Pat18} as they are an effective choice for transferring image appearances across domains. 
Semantic guidance is incorporated using a pre-trained semantic segmentation network trained on a different dataset.
 Semantic information in the form of segmentation feature maps from multiple levels is injected into the stain normalization network. Since we use self-supervised segmentation maps, we do not need manual annotations during training or test stages, which makes it easy to deploy for novel test cases. Our paper makes the following contributions:
 
  \begin{itemize}
  
  \item we integrate self-supervised features for stain normalization using semantic guidance from a pre-trained network;
  
  \item self-supervised segmentation feature maps allow us to use our method despite the unavailability of manual segmentation maps. Our proposed method beats the state-of-the-art stain normalization methods when the normalized images are used for classification and segmentation tasks. 

  \item Different from \cite{Lahiani,KuanarVC,MahapatraTMI2021,JuJbhi2020}, we explicitly use semantic information to capture geometric and structural patterns for image normalization. In particular, we use pixel adaptive convolution to match fine-grained segmentation maps of normalized images.

\end{itemize}

\section{Method}
\label{sec:method}

We denote the set of training images as $I_{Tr}$, their labels (manual segmentation masks or disease class) as $L_{Tr}$, and the trained model (segmentation or classification) as $M_{Tr}$. 
 Given a set of test images $I_{Test}$, our objective is to segment/classify them using the pre-trained model $M_{Tr}$. To successfully do that, we: 1) color normalize the test images using our proposed method SegCN-Net; and 2) apply pre-trained $M_{Tr}$. %

\begin{figure}[t]
 \centering
\begin{tabular}{c}
\includegraphics[height=4.7cm, width=11.5cm]{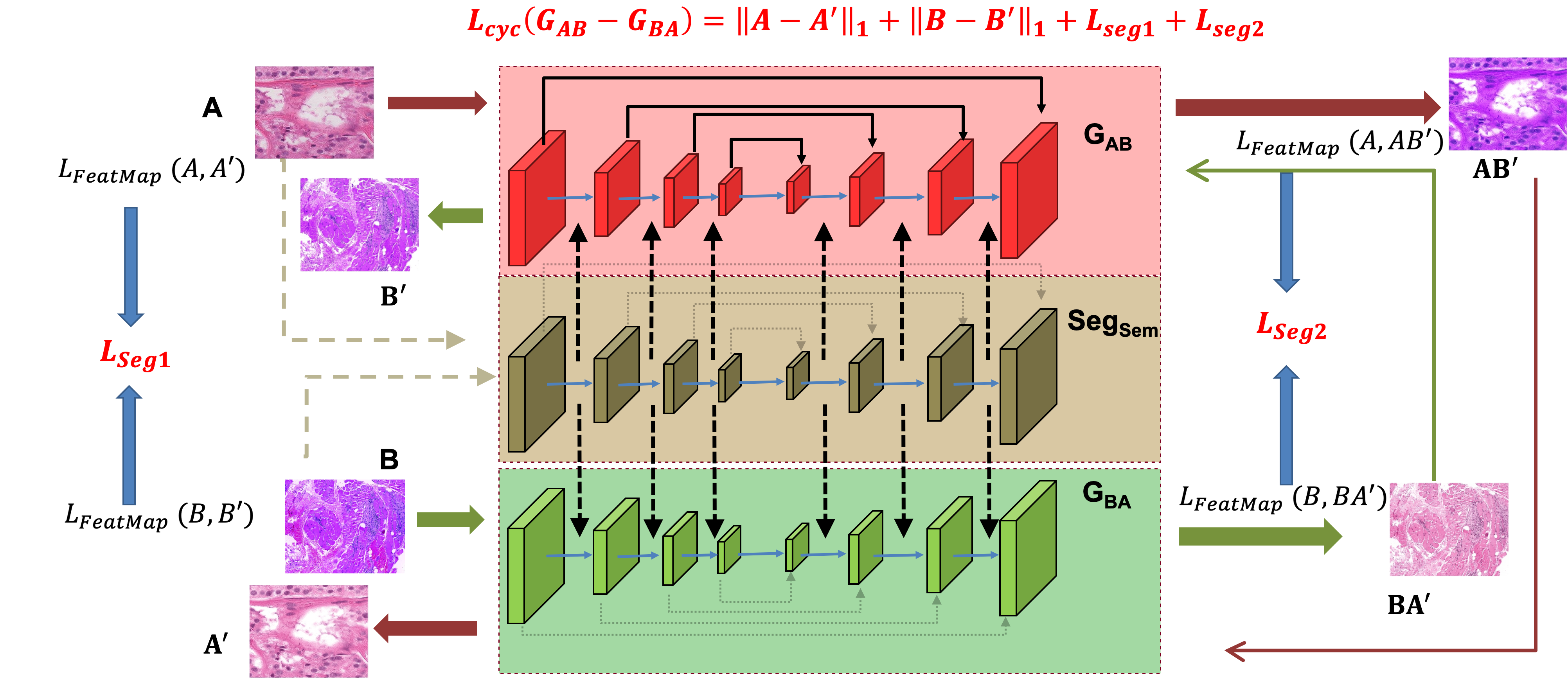}  \\
\end{tabular}
\caption{The workflow of our proposed stain color normalization method using CycleGANs. Semantic guidance is injected from the corresponding layers of $Seg_{Sem}$ into the generators $G_{AB},G_{BA}$, and helps to preserve critical cellular structures in normalization.}
\label{fig:overview}
\end{figure}

 Figure~\ref{fig:overview} depicts the workflow of our proposed stain normalization method. There are three different networks, $G_{AB}$ (the generator network in red), %  
 $Seg_{Sem}$ (the pre-trained segmentation network in yellow providing semantic guidance), and $G_{BA}$  (the generator network in green).% 
 All three networks are based on a UNet architecture \cite{Unet} to facilitate the easy integration of semantic information during training and test phases. 
 $G_{AB}$ transforms $ A $ to look like an image from domain $ B $ while $G_{BA}$ performs the reverse translation to maintain cycle consistency. Images from $A$ and $B$ are passed through $Seg_{Sem}$ and the information from different layers of $Seg_{Sem}$ is fused with the corresponding layer of $G_{AB}$ and $G_{BA}$ to facilitate the integration of semantic guidance.% 

\subsection{Semantic Guidance Through Self-Supervised Learning}

 Our self-supervised approach does not define any pretext task but focuses on using pre-trained networks for semantic guidance in stain normalization.  
 Semantic features for guiding the stain normalization task come from the pre-trained segmentation network $Seg_{Sem}$ shown in Figure~\ref{fig:overview}. $Seg_{Sem}$'s pre-trained weights guide the feature learning process of the two generators without the need for further finetuning. 

The translation invariance property of standard convolution makes it content-agnostic. It poses certain limitations, such as despite reducing the number of parameters, it may lead to sub-optimal learning of feature representations.
Additionally, spatially-shared filters globally average loss gradients over the entire image, and the learned weights can only encode location-specific information within their limited receptive fields. Content-agnostic filters find it difficult to distinguish between visually similar pixels of different regions (e.g., dark areas due to artifacts or tissues) or learn to identify similar objects of different appearances (e.g., same tissue structure with different shades as in our problem). % 

Pixel-adaptive convolutions \cite{SuCVPR19} can address the above limitations where the feature representations encoded in the semantic network help distinguish between confounding regions, and are defined as:
\begin{equation}
    \textbf{v}'_i=\sum_{j\in \Omega(i)} K(\textbf{f}_i,\textbf{f}_j) \textbf{W}\left[\textbf{p}_i-\textbf{p}_j\right] \textbf{v}_j +b
    \label{eqn:adconv}
\end{equation}
where $\textbf{f}$ are the features from the semantic network that guide the pixel adaptive convolutions, $\textbf{p}$ are pixel co-ordinates, $\textbf{W}$ is the convolutional weights of kernel size $k$, $\Omega_i$ is a $k\times k$ convolution window around pixel $i$, $\textbf{v}$ is the input and $b$ is the bias term. For each feature map, we apply a $3\times3$ and a $1\times1$ convolution layer followed by Group Normalization \cite{WuHe2018} and exponential linear units (ELU) non-linearities \cite{Clevert2016}. The resulting semantic feature maps are fused with the corresponding layers of $G_{AB}$ and $G_{BA}$ and used as guidance on their respective pixel-adaptive convolutional layers. $K$ is a standard Gaussian kernel defined by: 
\begin{equation}
     K(\textbf{f}_i,\textbf{f}_j)=\exp \left(-\frac{1}{2}(\textbf{f}_i-\textbf{f}_j)^{T} \Sigma^{-1}_{ij}(\textbf{f}_i-\textbf{f}_j) \right)
     \label{eqn:kernel}
\end{equation}
where $\Sigma^{-1}_{ij}$ is the covariance matrix between feature vectors $\textbf{f}_i$,$\textbf{f}_j$ and formulated as a diagonal matrix $\sigma^{2}\cdot I_D$, where $\sigma$ is an additional learnable parameter for each filter. %
 Standard convolution is a special case when $K(\textbf{f}_i,\textbf{f}_j)=1$.

To capture semantic information across multiple scales, we extract the feature maps after each convolution stage to get a set of maps with varying dimensions due to max pooling operations, whose values are normalized to $[0,1]$. For a given pair of images, we calculate the mean squared error between $f_s$ - corresponding multi-scale feature maps. Thus, the feature map loss between $a$ and $\hat{a}$ is:
\begin{equation}
 L_{FeatMap}(a,\hat{a})= \sum_{s=1}^{S} \sqrt{ \frac{\left(f_s(a)-f_s(\hat{a}) \right)^{2}} {N} }
\end{equation}

  \begin{figure}[t]
 \centering
\begin{tabular}{ccccc}
\includegraphics[height=2.3cm, width=2.3cm]{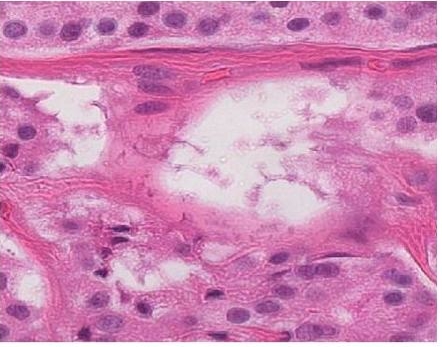} & 
\includegraphics[height=2.3cm, width=2.3cm]{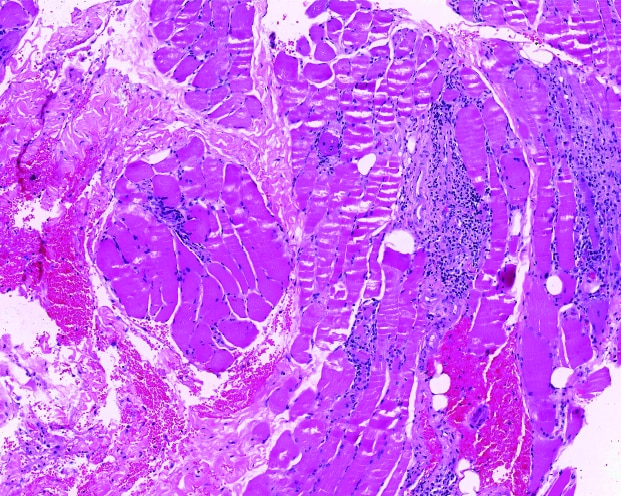} & 
\includegraphics[height=2.3cm, width=2.3cm]{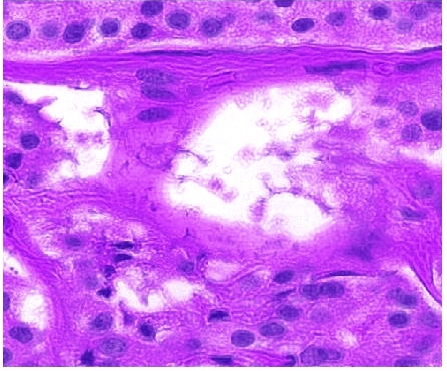} & 
\includegraphics[height=2.3cm, width=2.3cm]{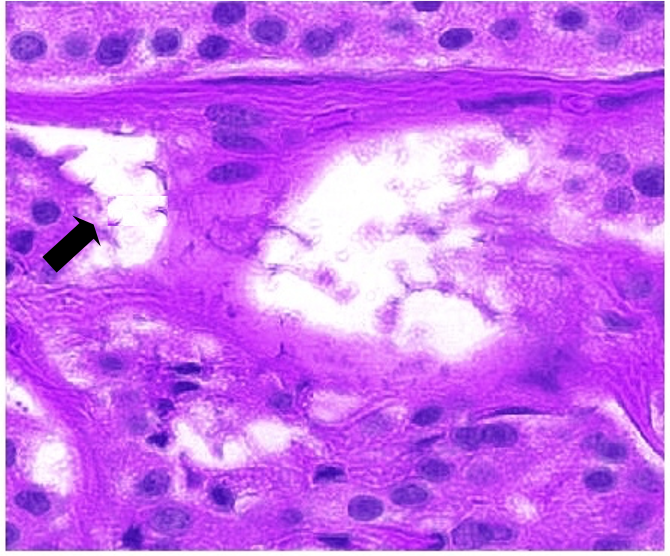} & 
\includegraphics[height=2.3cm, width=2.3cm]{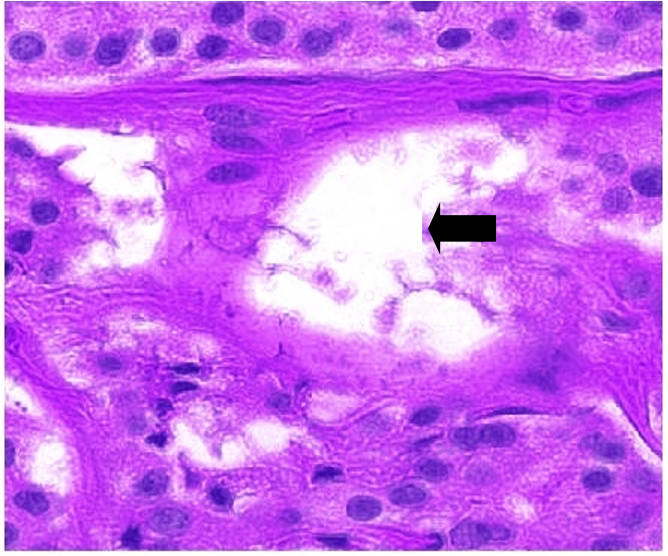} \\
(a) & (b) & (c) & (d) & (e) \\
\end{tabular}
\caption{Color normalization results: (a) Domain A image; (b) Domain B image;  Domain A transformed to Domain B using: (c) proposed $SegCN-Net$; (d) \cite{gadermayr2018miccai}; (e) \cite{ZhouMiccai19}. Areas of structure inconsistency are shown by black arrows.}
\label{fig:segRes}
\end{figure}

\subsection{Color Normalization Using Semantic Guidance}
Cycle GANs transform an image from domain $ A $ to $B$, and the reverse translation from $ B $ to $ A $ should generate the original input image. 
In forward cycle consistency, an image from domain $A$ is translated to domain $B$  by generator $G_{AB}$ expressed as $a\hat{b}=G_{AB}(a)$. Image $a\hat{b}$ is translated back to domain $A$ by $G_{BA}$ to get $\hat{a}=G_{BA}(a\hat{b})$.  Similarly, the original and reconstructed images from $B$ should also match.  Thus the overall cycle consistency loss is: % calculated as 
\begin{equation}
    L_{cycle} (G_{AB},G_{BA}) = E_{a} \left\|a -G_{BA}(G_{AB}(a))\right\|_1 + \left\|b -G_{AB}(G_{BA}(b))\right\|_1 +L_{Seg},  
    \label{eq:cyc}
\end{equation}
%
% where $L_{Seg}$ is defined as 
\begin{equation}
    L_{Seg} = \overbrace{L_{FeatMap}(a,\hat{a}) + L_{FeatMap}(b,\hat{b})}^{L_{Seg1}} + \overbrace{L_{FeatMap}(a,a\hat{b}) + L_{FeatMap}(b,b\hat{a})}^{L_{Seg2}}
    \label{eq:segL}
\end{equation}
We impose the additional constraint that the fine-grained segmentation maps of images should match, not just of the reverse transformed images $a,\hat{a}$ and $b,\hat{b}$) but also between the outputs of each generator and the corresponding original images, i.e. between $a,a\hat{b}$ and $b\hat{a},b$. $L_{Seg2}$ is specifically designed to preserve structural information between images of domain $A,B$ in stain normalization.

Discriminator $D_B$ is employed to distinguish between real image $b$
and generated image $a\hat{b}$ where the adversarial loss in forward cycle, $L_{adv}$, is:
%written as
\begin{equation}
    L_{adv} (G_{AB},D_{B},A,B) = E_{b} \log D_B (b) +  E_{a} \log \left[ \left(1-D_B (G_{AB}(a))\right)\right].
\end{equation}
There also exists a corresponding $L_{adv}(G_{BA},D_A,B,A)$ to distinguish between real image $a$ and generated image $b\hat{a}$
Thus the final objective function is: %optimized is 
\begin{equation}
    L= L_{adv}(G_{AB},D_B,A,B) + L_{adv}(G_{BA},D_A,B,A) + L_{cyc}(G_{AB},G_{BA})
\end{equation}

\subsubsection{Network Architecture}

All the networks ($G_{AB},G_{BA},Seg_{Sem}$) are based on a UNet architecture \cite{Unet} with a ResNet backbone. Each convolution block has $3$ layers of convolution layers (all using the adaptive pixel convolutions and ELU), followed by a $2\times2$ max-pooling step of stride $2$. Skip connections exist between the stages of the contracting and expanding path. $3\times3$ kernels are used with adequate padding to maintain image dimensions. There are four convolution blocks in both paths.

\section{Experimental Results}
\label{sec:expt}

\subsection{Evaluation Set Up for Classification}
\label{expt:desc}

Our proposed color normalization method is SegCN-Net (segmentation based color normalization network), and evaluate its performance as a pre-processing step. 
 CAMELYON16 \cite{CAMELYON16} and CAMELYON17 \cite{CAMELYON17} public datasets are used to have WSIs for classification and segmentation of breast cancer metastases. CAMELYON16 has images from $2$ independent medical centers, while CAMELYON17's images come from $5$ centers.
We train SegCN-Net on CAMELYON16 and evaluate on transformed images of CAMELYON17. Domain $A$ consists of images from Center~1 of CAMELYON16 ($C1_{16}$), while domain $B$ has images from $C2_{16}$. 
$100,000$ patches of $256\times256$  were extracted from each domain, and we train all models using an NVIDIA Titan X GPU having $12$ GB RAM, Adam optimizer \cite{Adam} with a learning rate of $0.002$.
Xavier initialization was used, and training took $42$ hours for $150$ epochs with batch size $16$.

For evaluation, images from the different centers of CAMELYON17 were split into training/validation/test in  $50/30/20 \%$ to obtain the following split: $C1_{17}$:37/22/15, $C2_{17}$: 34/20/14, $C3_{17}$: 43/24/18, $C4_{17}$: 35/20/15, $C5_{17}$: 36/20/15. For our first baseline, we train $5$ different  ResNet50 \cite{ResNet} with batch size $32$, Adam optimizer learning rate of $0.001$ for $70$ epochs (denoted as \textbf{ResNet$_{NoNorm}$}) on images from $C1_{17} - C5_{17}$ using the split described before, but without normalization.  
We apply SegCN-Net on images from different centers of CAMELYON17 to color normalize them and train ResNet50 networks with similar settings as $ResNet_{NoNorm}$ using the data split of $C1_{17} - C2_{17}$. The results (using the area under the curve (AUC) as the performance metric) are reported in Table~\ref{tab:NormClass} under SegCN-Net.
We replace our stain normalization method with other competing methods, such as 
\cite{ZhouMiccai19,Reinhard,Machenko,Vahadane,gadermayr2018miccai} and perform the same set of classification experiments
with the performance summarized in Table~\ref{tab:NormClass}.

\cite{Vahadane} aims to preserve structure information through templates while \cite{ZhouMiccai19} employ
stain color matrix matching. Since they do not explicitly use segmentation information, SegCN-Net performs better than both methods. The method by \cite{gadermayr2018miccai} does better than others because of the use of segmentation information but requires labeled segmentation maps. SegCN-Net's superior performance shows that the use of self-supervised segmentation can be leveraged when manual segmentation maps are not available. 
Figure~\ref{fig:segRes} shows the stain normalized images of different methods. 
The advantage of SegCN-Net in preserving  structural information is indicated by the 
black arrows where the glandular structure is deformed from the original image in 
\cite{ZhouMiccai19}, and to a lesser extent in \cite{gadermayr2018miccai}. 
Thus the advantages of our semantic guidance based stain normalization are obvious.

\begin{table}[t]
 \begin{center}
 \caption{Classification results in terms of AUC measures for different stain normalization methods on the CAMELYON17 dataset. $p$ values are with respect to SegCN-Net.}
\begin{tabular}{|c|c|c|c|c|c|c|c|}
\hline 
{Method} & {$Center~1$}  & {$Center~2$}  & {$Center~3$} & {$Center~4$} & {$Center~5$} & {$Average$} & {$p$}\\ \hline
{$ResNet_{C17noNorm}$}& {0.8068}  & {0.7203} & {0.7027} & {0.8289} & {0.8203} & {0.7758}  & {0.0001} \\ \hline
{Reinhard \cite{Reinhard}} & {0.7724} & {0.7934} & {0.8041} & {0.8013} & {0.7862} & {0.7915} & {0.0001} \\ \hline
{Macenko \cite{Machenko}} & {0.7148} & {0.7405}  & {0.8331} & {0.7412} & {0.7436} & {0.7546} & {0.0001} \\ \hline
{CycleGAN} & {0.9010} & {0.7173}  & {0.8914} & {0.8811} & {0.8102} & {0.8402} & {0.002} \\ \hline
{Vahadane \cite{Vahadane}} & {0.9123} & {0.7347} & {0.9063} & {0.8949} & {0.8223} & {0.8541} & {0.003} \\ \hline
{Zhou \cite{ZhouMiccai19}} & {0.9381} & {0.7614} & {0.7932} & {0.9013} & {0.9227} & {0.8633} & {0.013}\\ \hline
{Gadermayr \cite{gadermayr2018miccai}} & {0.9487} & {0.8115} & {0.8727} & {0.9235} & {0.9351} & {0.8983} & {0.013}\\ \hline
{SegCN-Net} & {\textbf{0.9668}} & {\textbf{0.8537}}  & {\textbf{0.9385}} & {\textbf{0.9548}} & {\textbf{0.9462}} & {\textbf{0.9320}} & {-} \\ \hline
\multicolumn{8}{c}{\textbf{Ablation Study Results}} \\ \hline
{SegCN-Net$_{Conv}$} & {0.9331}  & {0.8255} & {0.9148} & {0.9259} & {0.9181} & {.9035} & {0.0008} \\ \hline
{SegCN-Net$_{Seg~Only}$} & {0.9376} & {0.7974}  & {0.8942} & {0.9187} & {0.9012} & {0.8898}  & {0.0001} \\ \hline
{SegCN-Net$_{C17Rand}$} & {0.9624} & {0.8403}  & {0.9267} & {0.9478} & {0.9391} & {0.9232} & {0.34} \\ \hline
{SegCN-Net$_{Glas}$} & {0.9762} & {0.8627}  & {0.9509} & {0.9677} & {0.9588} & {0.9432} & {0.042} \\ \hline
\end{tabular}
\label{tab:NormClass}
\end{center}
\end{table}

\subsection{Ablation Studies}

Table~\ref{tab:NormClass} summarizes the performance of the following variants of our method:
\begin{enumerate}
    \item SegCN-Net$_{Conv}$ - $SegCN-Net$ using standard convolutions instead of pixel adaptive convolutions.
    \item SegCN-Net$_{Seg}$ - SegCN-Net using only the final segmentation masks without the intermediate feature map. This evaluates the relevance of using a single segmentation map without semantic guidance at each layer.
    \item SegCN-Net$_{C17Rand}$ - SegCN-Net tested on all normalized images of  C17 with a random selection of train/val/split. The results are an average of $10$ runs and investigate possible bias in data split.
\end{enumerate}

In the original approach, $Seg_{Sem}$ was pre-trained on the MS-COCO dataset \cite{MS-COCO}. In a variant of our proposed method, we use a network pre-trained on the Glas segmentation challenge dataset \cite{GlasReview}, which has segmentation masks of histological images and use it for classification of the test images from CAMELYON17. The results are shown in Table~\ref{tab:NormClass} under SegCN-Net$_{Glas}$. 

SegCN-Net$_{Glas}$ shows better classification performance than 
SegCN-Net and the difference in results at $p=0.042$ is significant  as semantic guidance is obtained from a network
trained on histology images, while SegCN-Net used natural images.
Although natural images provide some degree of semantic guidance by learning edge features,
$Seg_{Sem}$ trained on histopathology images provides domain-specific guidance and hence leads to better performance. Since such an annotated dataset is not always available for medical images, we show that semantic guidance from a network trained on natural images significantly improves the state-of-art method for stain color normalization. % 

SegCN-Net$_{C17Rand}$ performance is close to SegCN-Net without any statistically significant 
the difference, indicating that SegCN-Net is not biased on the test set. 
SegCN-Net$_{Seg~Only}$ shows inferior performance compared to SegCN-Net, which indicates
that multistage semantic guidance is much better than a single segmentation map. However 
SegCN-Net$_{Seg~Only}$ still performs slightly better than \cite{gadermayr2018miccai} indicating the advantages of including segmentation information for structure-preserving color normalization.

\subsection{Segmentation Results}

We apply our method on the public GLAS segmentation challenge \cite{GlasReview}, which has manual segmentation maps of glands in $165$ $H\&E$ stained images derived from $16$ histological sections from different patients with stage $T3$ or $T4$ colorectal adenocarcinoma. %
We normalize the images using SegCN-Net (using MS-COCO images for semantic guidance), train a UNet with residual convolution blocks, and apply on the test set. The performance metrics - Dice Metric (DM), Hausdorff distance (HD), F1 score (F1)- for SegCN-Net, \cite{ZhouMiccai19,gadermayr2018miccai} and the top-ranked method \cite{GlasResults} are summarized in Table~\ref{tab:seg}. \cite{ZhouMiccai19}'s performance comes close to the top-ranked while \cite{gadermayr2018miccai} outperforms both of them, and SegCN-Net gives the best results across all three metrics. This shows that stain normalization, in general, does a good job 
of standardizing image appearance, which in turn improves segmentation results.
SegCN-Net performs best due to the integration of segmentation information through self-supervised semantic guidance.

\begin{table}[t]
 \begin{center}
 \caption{Segmentation results on the GLas Segmentation challenge for $SegCN-Net$, \cite{gadermayr2018miccai}, \cite{ZhouMiccai19} and the top ranked method. $HD$ is in mm. Best results per metric in bold.}% HD is in mm}
\begin{tabular}{|c|c|c|c|c|c|c|c|c|}
\hline 
{} & \multicolumn{2}{|c|}{$SegCN-Net$}  & \multicolumn{2}{|c|}{Glas Rank 1}  & \multicolumn{2}{|c|}{\cite{gadermayr2018miccai}} & \multicolumn{2}{|c|}{\cite{ZhouMiccai19}} \\ \hline
{} & {Part A} & {Part B} & {Part A} & {Part B} & {Part A} & {Part B} & {Part A} & {Part B} \\ \hline
{F1} & {\textbf{0.9351}} & {\textbf{0.7542}} & {0.912} & {0.716} & {0.926} & {0.728} & {0.922} & {0.729} \\ \hline
{DM} & {\textbf{0.9212}} & {\textbf{0.8054}} & {0.897} & {0.781} & {0.909} & {0.798} & {0.892} & {0.785} \\ \hline
{HD} & {\textbf{42.276}} & {\textbf{143.286}} & {45.418} & {160.347} & {44.243} & {157.643} & {47.012} & {161.321} \\ \hline
\end{tabular}
\label{tab:seg}
\end{center}
\end{table}

\subsection{Color Constancy Results} 

Similar to \cite{Zanjani} we report results for normalized median intensity, which measures color constancy of images, for the same dataset and obtained the following values: SegCN-Net - Standard Deviation(SD)$=0.011$, Coefficient of Variation (CV)$=0.021$, which is better than Zanjani et al. \cite{Zanjani} - $SD=0.0188, CV=0.0209$.

As reported in \cite{Lahiani}, we calculate values for complex wavelet structural similarity index (CWSSIM) between real and generated images. CWSSIM $\in[0,1]$ with higher values indicating a better match and is robust to small translations and rotations. Mean CWSSIM values of SegCN-Net is $0.82$, which is higher than CycleGAN ($0.75$) and \cite{Lahiani} ($0.77$).

% this file has the conclusion for the paper

\section{Conclusion}
\label{sec:concl}
We have proposed a novel approach to stain-color normalization in histopathology images, integrating semantic features into the self-supervised network. Our semantic guidance approach facilitates the inclusion of segmentation information without the need for manually segmented maps that are very difficult to obtain. Experimental results on public datasets show our approach outperforms the state of the art normalization methods when evaluated for classification and segmentation tasks. Ablation studies also show the importance of semantic guidance. Although semantic guidance is obtained from a held-out dataset, we also demonstrate that when domain-specific guidance is used, the results improve even further. This has the potential for enhancing the performance of medical image analysis tasks where annotations are not readily available.

%\tikz[overlay,remember picture]
%    \draw (current page.west) -- (current page.east);
    
%We have proposed a histopathology image stain color normalization approach using cycle GANs that integrates semantic guidance from self supervised segmentation feature maps. Our semantic guidance approach facilitates inclusion of segmentation information without the need for manually segmented maps that are very difficult to obtain. Experimental results on public datasets show our approach outperforms state of the art normalization methods when evaluated for  classification and segmentation. Ablation studies also show the importance of semantic guidance. Although semantic guidance is obtained from the MS-COCO dataset of natural images, we also demonstrate that when domain specific guidance is used the results improve even further. This has potential in improving performance of medical image analysis tasks where annotations are not readily available.  

%
 \bibliographystyle{splncs04}
 \bibliography{MICCAI2020_StainNorm,MyCitations_Conf,MyCitations_Journ}

\end{document}